\documentclass[fleqn,usenatbib]{mnras}

\usepackage{newtxtext,newtxmath}

\usepackage[T1]{fontenc}

\DeclareRobustCommand{\VAN}[3]{#2}
\let\VANthebibliography\thebibliography
\def\thebibliography{\DeclareRobustCommand{\VAN}[3]{##3}\VANthebibliography}

\usepackage{graphicx}	
\usepackage{amsmath}	
\newcommand\cmc{{\rm CMC}}

\title[Photodynamic agent in TLCM]{Modelling the Light Curves of Transiting Exomoons: a Zero-order Photodynamic Agent Added to the Transit and Light Curve Modeller}

\author[Sz. K\'alm\'an]{
Szil\'ard K\'alm\'an,$^{1,2,3}$\thanks{E-mail: xilard1@gothard.hu (SzK)}
Szil\'ard Csizmadia,$^{4,5}$
Attila Simon E.,$^{6,7}$
Kristine W. F. Lam,$^{4}$
Adrien Deline,$^{8}$ \newauthor Jan-Vincent Harre,$^{4}$
and Gyula Szab\'o M.$^{3,9}$
\\
$^{1}$ HUN-REN, Research Centre for Astronomy and Earth Sciences, Konkoly Observatory, MTA Centre of Excellence, \\H-1121 Budapest, Konkoly Thege Mikl\'os \'ut 15-17, Hungary\\
$^{2}$ELTE E{\"o}tv{\"o}s Lor\'and University
Doctoral School of Physics, Pázmány Péter sétány 1/A,  Budapest, H-1117, Hungary\\
$^{3}$HUN-REN–ELTE Exoplanet Systems Research Group, Szent Imre h. u. 112., Szombathely, H-9700, Hungary\\
$^4$Deutsches Zentrum für Luft- und Raumfahrt, Institute of Planetary Research, Rutherfordstrasse 2, D-12489 Berlin, Germany \\ 
$^5$HUN-REN–SZTE Stellar Astrophysics Research Group, H-6500 Baja, Szegedi \'ut Kt. 766, Hungary\\
$^6$Physikalisches Institut, University of Bern, Gesellschaftsstrasse 6, 3012 Bern, Switzerland\\
$^7$Center for Space and Habitability, University of Bern, Gesellschaftsstrasse 6, 3012 Bern, Switzerland\\
$^8$Department of Astronomy, University of Geneva, Chemin Pegasi 51, 1290 Versoix, Switzerland\\
$^{9}$ELTE E{\"o}tv{\"o}s Lor\'and University, Gothard Astrophysical Observatory, Szent Imre h. u. 112., H-9700, Hungary
}

\date{Accepted 2023 November 07. Received 2023 November 07; in original form 2023 August 31}

\pubyear{2023}

\begin{document}
\label{firstpage}
\pagerange{\pageref{firstpage}--\pageref{lastpage}}
\maketitle

\begin{abstract}
Despite the ever-growing number of exoplanets discovered and the extensive analyses carried out to find their potential satellites, only two exomoon candidates, Kepler-1625b-i and Kepler-1708 b-i, have been discovered to date. A considerable amount of effort has been invested in the development of algorithms for modelling, searching, and detecting exomoons in exoplanetary light curves. 
In this work, we incorporate moon handling capabilities into the state-of-the-art and publicly available code, the Transit and Light Curve Modeller (TLCM).
The code is designed for the analysis of transiting exoplanet systems with the inclusion of a wavelet-based noise handling algorithm. Here we present an updated version of TLCM that is capable of modelling a coplanar planet-moon system on an elliptical orbit around its host, accounting for mutual eclipses between the two bodies (and neglecting perturbative effects) -- a so-called photodynamic model. The key benefit of this framework is the ability for a joint analysis of multiple planet--moon transits. We demonstrate the necessity of this software on a case study of Kepler-1625b. Similarly to prior works, we conclude that there is no firm evidence of an exomoon in that system, by showing that temporally correlated noise can mimic apparent lunar transits.
\end{abstract}

\begin{keywords}
celestial mechanics -- techniques: photometric -- planets and satellites: detection -- planets and satellites: individual: Kepler-1625b
\end{keywords}



\section{Introduction}

Since the discovery of the first extrasolar planet orbiting main sequence stars \citep[e.g.][]{1995Natur.378..355M}, the field of exoplanet research grew rapidly and became increasingly important in astronomy. Based on our understanding of the Solar System, the quest to find satellites of these planets (called `exomoons') began to expand shortly thereafter \citep[e.g.][]{1999A&AS..134..553S}. Although there are $>5000$ known exoplanets to date, only two exomoon-candidates have been identified \citep[Kepler-1625b-i;][]{2018SciA....4.1784T} and \citep[Kepler-1708 b-i;][]{2022NatAs...6..367K}. \cite{2019A&A...624A..95H} offer an alternative no-moon explanation for the signal observed by \cite{2018SciA....4.1784T}, involving the presence of time-correlated noise  and the properties of Bayesian inference. In \cite{2019ApJ...877L..15K}, the authors also argue that the there is no conclusive evidence for an exomoon in the transit light curves of Kepler-1625b. One can therefore argue that the transiting exomoon candidate proposed by \cite{2018SciA....4.1784T} is the result of temporally variable noise mimicking the transit of a moon. There are two distinct sources of this noise type (also commonly called red noise): astrophysical and instrumental. The former of these include the presence of stellar pulsations, flares, granulation, and spots, which may produce light variations similar to the ones expected from transits of solid bodies \citep[e.g.][]{2010A&A...510A..25S}. The latter group includes cosmic rays causing excess flux, temperature variations of the detectors/telescope assembly, pointing drifts etc \citep[see e.g.][for more discussion on this topic]{2021arXiv210811822C}. Time-correlated noise can manifest on a wide range of timescales. The long-term trends can easily be accounted for by linear or parabolic baselines \citep{2018SciA....4.1784T}. \cite{2019ApJ...877L..15K} suggests that instrumental noise sources (of shorter time-scales) may be behind the apparent exomoon transit. 

Furthermore, \cite{2021MNRAS.501.2378F} found six exomoon candidates (KOIs-268.01, 303.01, 1888.01, 1925.01, 2728.01, 3320.01) based on Transit Timing Variations (TTVs). \cite{2020ApJ...900L..44K} disproved these specific exomoon candidates from observational aspects, while \cite{2020ApJ...902L..20Q} suggest that the systems mentioned above could not be stable from a dynamical perspective. The field of exomoon research is intertwined with current and upcoming space-based telescopes, such as CHEOPS \citep{2015PASP..127.1084S, 2021ExA....51..109B, 2023A&A...671A.154E}, PLATO \citep{2014ExA....38..249R} and Ariel \citep{2021arXiv210404824T, 2022ExA....53..607S}, the number of potentially discoverable exomoons is therefore expected to rise. 

In recent decades, a number of methods were proposed to detect and characterise exomoons. These include, but are not limited to, the analysis of TTVs \citep{2007A&A...470..727S, 2009MNRAS.392..181K}, the Rossiter-McLaughlin effect \citep{2010MNRAS.406.2038S}, direct modelling of the transits \citep[e.g.][]{2011ESS.....2.1902K}, finding radio signatures caused by the interaction of the moon and its host planet \citep[e.g.][]{2014ApJ...791...25N, 2016ApJ...821...97N}, spectrophotometry \citep{2021MNRAS.508.5524J} and direct imaging \citep{2023A&A...675A..57K}. As expressed above, some of these methods were used to identify exomoon candidates. The long-term stability of the planet-moon systems is also favoured from an observational point of view. A considerable effort was made to assess the dynamics of the formation and evolution of such systems in general \citep{Domingos2006,2016ApJ...817...18S,2017MNRAS.471.3019A,2019MNRAS.489.2313S}, and the satellites of Kepler-1625b/Kepler-1708b \citep{2020MNRAS.495.3763M, 2023MNRAS.520.2163M}. The topic of habitability is also often discussed in relation to the exomoons \citep[e.g.][]{2009MNRAS.400..398K, 2010ApJ...712L.125K, 2022MNRAS.513.5290D}.

Following the success of the transit method in the detection of exoplanets, several algorithms have been put forward to model the light curve of a transiting planet-moon system \citep[e.g.][]{2017ApJ...851...94L,2021MNRAS.507.4120K, 2022ApJ...936....2S, 2022AJ....164..111G}. At the time of writing however, there is only one open-source software available that is capable of fitting the light curves of transiting planet-moon systems out-of-the-box \citep{2022A&A...662A..37H}. Recently, time-correlated noise (either of stellar or instrumental origin) has been demonstrated to mimic exomoon light curve components, and is recognised to be a major source of false alarms \citep[e.g.][]{2023A&A...671A.154E}. Therefore, there is a clear need for algorithms with a combination of noise-handling capabilities and the photodynamic effects. Red noise can arise from instrumental and astrophysical sources, such as stellar spots, microflares, pulsation etc. \citep[see][for more detailed discussions]{2021arXiv210811822C, 2022arXiv220801716K}. To filter out the false-positive moon signals that are consistent with noise level fluctuations, a light curve inversion algorithm that accounts for red noise is needed. 

In this paper, we introduce the photodynamic update of the Transit and Light Curve Modeller \citep[TLCM;][]{2020MNRAS.496.4442C}, which uses the well-established wavelet-based noise filtering algorithm of \cite{2009ApJ...704...51C}. Its benefits in the precision and accuracy of exoplanet parameters in transit light curves are well established \citep{2021arXiv210811822C, 2022arXiv220801716K}. The newly integrated model also includes the ability to model mutual transits by the planet and its satellite. In Section \ref{sec:meth}, we describe the calculation of the position of the planet and its satellite. In Section \ref{sec:LCs}, we provide example light curves with arbitrary parameters and a specific case of Kepler-1625b-i \citep{2018SciA....4.1784T}. In Section \ref{sec:con}, we draw our conclusions.

\section{Methods} \label{sec:meth}

\begin{figure*}
    \centering
    \includegraphics[width = \textwidth]{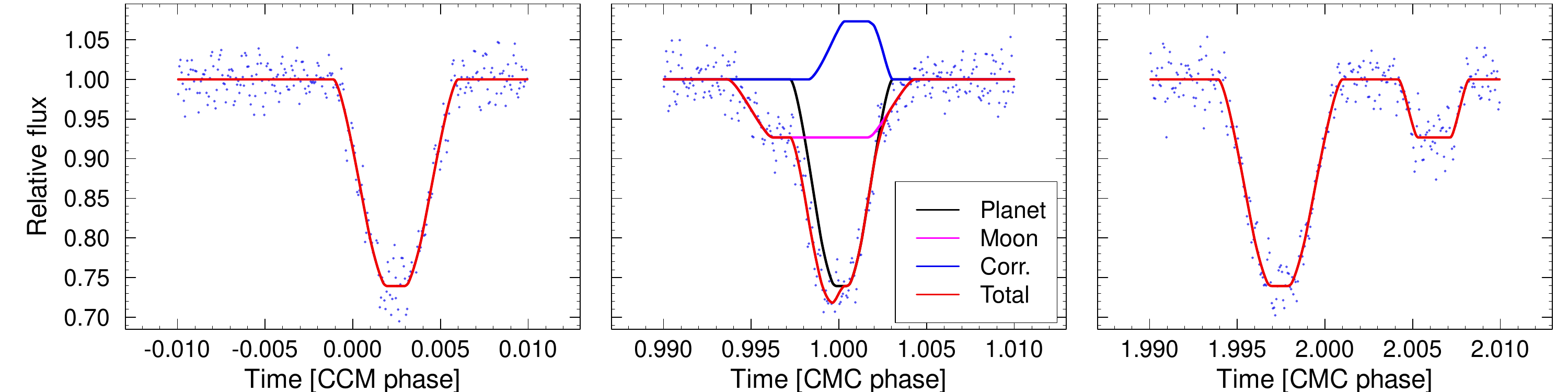}
    \caption{Simulated light curve of an arbitrary transiting planet-moon system (blue dots and solid red curve). \textbf{Left.} The moon remains hidden behind the planet during the transit. \textbf{Middle.} A mutual transit event occurs, requiring flux correction (solid blue curve). \textbf{Right.} The transit of the satellite occurs after the transit of the exoplanet. Note that, for demonstrative purposes, the sizes of both bodies are enlarged, and we assume no limb darkening of the host star.}
    \label{fig:demo}
\end{figure*}

The TLCM software is designed to model a wide range of astrophysical effects that can be observed in light curves of transiting exoplanets, including transits, occultations, phase curve variability, ellipsoidal effect and Doppler beaming \citep{2020MNRAS.496.4442C, 2021arXiv210811822C}. Special emphasis is given to the handling of time-correlated noise: the wavelet-based noise filtering of \cite{2009ApJ...704...51C} is used to remove (or reduce) the amount of red noise present in light curves \citep{2021arXiv210811822C, 2022arXiv220801716K}. 
This noise filtering maximises the likelihood of the combination of the noise and transit models, thus `forcing' the signal into the shape with the highest likelihood.

To add the transits of exomoons to TLCM, we make the following assumptions. The common center of mass (CCM) of the planet-moon system has an elliptical motion around the host star, and both of them are on circular orbits around the CCM. Furthermore, the planet and its satellite orbit the CCM in the same plane as the CCM orbits the star.

These assumptions help us avoid high degrees of degeneracy \citep{2010MNRAS.406.2038S,2022A&A...662A..37H}, and to limit the number of free parameters to five to describe the moon \citep{2002ApJ...580L.171M, 2020MNRAS.496.4442C}. These are the scaled semi-major axis of the moon, $a_{\rm Moon}/R_{\rm Planet}$, the moon's relative radius $R_{\rm Moon}/R_{\rm Planet}$, its relative mass, $q = M_{\rm Moon}/M_{\rm Planet}$, its orbital period $P_{\rm Moon}$, and $\Delta \varphi$ -- a phase shift between the epoch of the CCM and the transit of the satellite. The complete transit light curves are computed via numerical integration of the two bodies, whose sky-projected positions are calculated via the equations described below. The model presented here is also capable of accounting for mutual eclipses of the planet and its satellite, as shown on Fig. \ref{fig:demo}. 

We describe these orbits by introducing two phase angles:
\begin{align}
    \varphi_\cmc &= 2 \pi \frac{t-t_0}{P_\cmc} \\
    \varphi_{\rm Moon} &= 2 \pi \frac{t-t_0}{P_{\rm Moon}},
\end{align}
where $t_0$ is the epoch at which the CCM is collinear with the center of the stellar disk and the observer, $P_\cmc$ and $P_{\rm Moon}$ are the orbital periods of the CCM around the star and of the  satellite around the planet, respectively. The mean anomaly at $t_0$, denoted by $M_0$, is calculated via Eqs. (12-15) of \citet{2020MNRAS.496.4442C}. The mean anomaly can then be written as $M = \varphi_\cmc + M_0$. Let $e_\cmc$ and $\omega_\cmc$ denote the eccentricity and the argument of the periastron (measured from the tangential plane of the sky) of the CCM, respectively. Using Kepler's equation, the eccentric anomaly, solved iteratively, can be written as $E_\cmc = M + e_\cmc \sin E_\cmc$. The true anomaly, $\nu$, can then be calculated as
\begin{equation}
    \tan \frac{\nu_\cmc}{2} = \sqrt{\frac{1+e_\cmc}{1-e_\cmc}} \tan \frac{E_\cmc}{2}.
\end{equation}

In the orbital plane of the planet, the position of the common center of mass at any point in time is
\begin{align}
    x_\cmc  &= \frac{a_\cmc}{R_\star} \cos \left( \nu_\cmc + \omega_\cmc  \right) \frac{1-e^2_\cmc}{1+e \cos \left( \nu_\cmc \right)} \\
    y_\cmc &= \frac{a_\cmc}{R_\star} \sin \left( \nu_\cmc + \omega_\cmc \right) \frac{1-e^2_\cmc}{1+e\cos\left(\nu_\cmc \right)} \\
    z_\cmc &= 0,
\end{align}
where $a_\cmc$ is the semi-major axis of the CCM, and $R_\star$ is the stellar radius. With the introduction of the moon-to-planet mass ratio, $q = M_{\rm Moon}/M_{\rm Planet}$, and the semi-major axis of the orbit of the moon, $a_{\rm Moon}$, the position of the planet in the orbital plane can be written as
\begin{align}
    x_{\rm Planet} &= x_\cmc - \frac{a_{\rm Moon}}{R_{\rm Planet}} \frac{R_{\rm Planet}}{R_\star} \frac{q}{1+q} \cos \left( \varphi_{\rm Moon} + \Delta_\varphi \right) \label{eq:xp} \\
    y_{\rm Planet} &= y_\cmc - \frac{a_{\rm Moon}}{R_{\rm Planet}} \frac{R_{\rm Planet}}{R_\star} \frac{q}{1+q} \sin \left( \varphi_{\rm Moon} + \Delta_\varphi \right) \\
    z_{\rm Planet} &=0.
\end{align}
The position of the moon in the orbital planet of the planet is given by
\begin{align}
    x_{\rm Moon} &= x_\cmc + \frac{a_{\rm Moon}}{R_{\rm Planet}} \frac{R_{\rm Planet}}{R_\star} \frac{1}{1+q} \cos \left( \varphi_{\rm Moon} + \Delta_\varphi \right) \\
    y_{\rm Moon} &= y_\cmc + \frac{a_{\rm Moon}}{R_{\rm Planet}} \frac{R_{\rm Planet}}{R_\star} \frac{1}{1+q} \sin \left( \varphi_{\rm Moon} + \Delta_\varphi \right) \\
    z_{\rm Moon} &=0. \label{eq:zm}
\end{align}
In the equations above, $\Delta \varphi$ is the phase shift of the epoch of the satellite, and $R_{\rm Planet}$ is the planetary radius. The sky-projected position of the exomoon and the exoplanet are then calculated from Eqs. (\ref{eq:xp} -- \ref{eq:zm}) via the equations described in Sect. 2.2. of \cite{2020MNRAS.496.4442C}.

\begin{figure*}
    \centering
    \includegraphics[width = \textwidth]{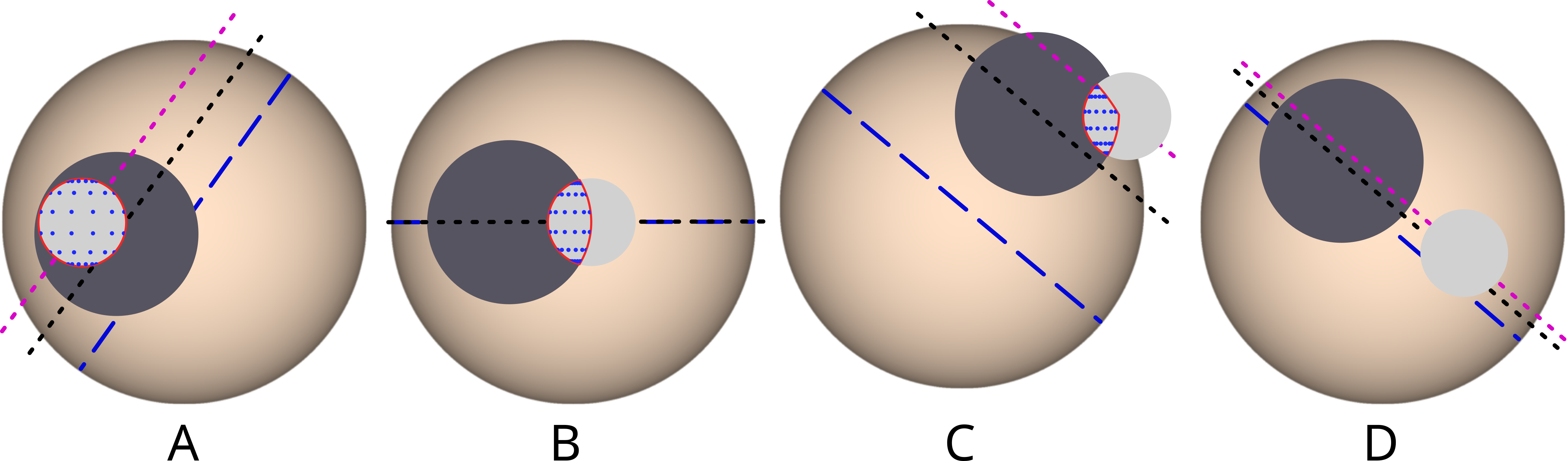}
    \caption{Possible mutual eclipse cases with an illustration of the Gauss-Legendre quadrature integration points (blue) distributed during the mutual eclipses. The three discs represent, in decreasing order of diameter, the star, the planet and the moon. The projected orbits of the planet and the moon are shown with dashed black and magenta lines. The dashed blue line is used to visualise the impact parameters of the two bodies. The limb darkening of the stellar disc is generated via solar-like parameters according to the routines of \protect\cite{2021AN....342..578H}. The mutually eclipsed area (by the planet and the moon) is outlined with solid red lines. \textbf{Case A}: the planet fully occults the moon (or the moon totally transits the planet, i.e. the moon can be behind or in front of the planet). \textbf{Case B}: The planet and the moon exhibit partial mutual eclipse but both totally eclipses the star.\textbf{ Case C}: During the partial eclipses of the planet and/or the moon, three arcs may be needed to find the mutually eclipsed area. \textbf{Case D}: no mutual eclipse.}
    \label{fig:eclipses}
\end{figure*}

From the known sky-projected positions, the limb-darkened transits of the two bodies are calculated via the commonly used numerical integration tools to account for the light blocked by the planet and the moon. The resulting light curve therefore consists a `planet component' and a `moon component' superimposed on each other \citep{2007A&A...470..727S}. Owing to the assumed coplanarity, the projected orbital trajectories of the planet and the moon will only coincide if the inclination of the orbital plane of the CCM (with respect to the line of sight) is $90^\circ$. This is highligted on Fig. \ref{fig:eclipses}.

Furthermore, there is a check for the occurrence of mutual transits (i.e. the moon passing in front of, or behind the planet). If such an event is detected, a correction is applied so that the mutually eclipsed area is calculated only once. A mutual transit event can occur if and only if all of the following conditions are fulfilled:
\begin{align}
    \delta_\mathrm{Planet-Moon} &\leq R_\mathrm{Planet} + R_\mathrm{Moon} \\
    \delta_{\star \mathrm{-Planet}} &\leq R_\star + R_\mathrm{Planet} \\
    \delta_{\star \mathrm{-Moon}} & \leq R_\star + R_\mathrm{Moon} 
\end{align}

%
%
%
%
%
%
Here $\delta$ denotes the sky-projected mutual distance of the different objects. When all three conditions are simultaneously fulfilled, the arcs of the circular boundaries between the star, planet and moon are determined to calculate the mutually eclipsed area. This is carried out in a numerical way. A 101x101 pixel image is centered around the moon and planet and we check which points are inside the sky-projected images of the star, planet and moon. The mutually eclipsed area is enclosed by either a full circle (in the limit if the moon is completely behind or in front of the planet, or, conversely if the planet is smaller than the moon), or by two or three arcs (see Fig. \ref{fig:eclipses}).
The user can choose between $6\times6$, $10\times10$, $20\times20$, $48\times48$ or $96\times96$ integration points which are distributed over the mutually eclipsed area according to the Gauss-Legendre quadrature rules. According to our experiences, a $6\times6$ grid is enough to get a few ppm precision. The intensity behind the mutually eclipsed area is then calculated with the Gauss-Legendre quadrature and the limb darkening is taken into account. This correction flux is added accordingly to the light curve, because the simple transit events of the moon and the planet would result in a duplicated subtraction of the flux behind the mutually eclipsed area. An illustration of how this works is given by Figs. \ref{fig:demo} and \ref{fig:eclipses}. 

We note that the current implementation of the photodynamic model is highly simplified, as the orbital plane of the moon is considered to coincide with the one of the CCM. Furthermore, the two bodies are also considered to be on Keplerian orbits (hence the term `zero order'). Future updates will focus on the inclusion of the possibility of misaligned orbital plane for the satellite, as well as the inclusion of various perturbative effects, which may lead to variations in periodicity, transit depth and duration etc. \citep[see e.g.][for a more detalied discussion on these effects in triple star systems]{2022MNRAS.510.1352B}.

\section{Light curve simulation} \label{sec:LCs}

\subsection{A schematic configuration}
An example light curve, computed via the methods described in Sect. \ref{sec:meth}, is shown on Figure \ref{fig:demo}. This system was set up with the following exaggerated, arbitrary parameters: $a_\cmc/R_\star = 50$, $a_{\rm Moon}/R_{\rm Planet} = 10.2$, $\Delta \varphi = 0.4$, $R_{\rm Moon}/R_{\rm Planet} = 0.53$, $P_{\rm Moon} = 2.5789$ days, $R_{\rm Planet}/R_\star = 0.51$, $q = 0.3$, $e_\cmc = 0.35$,  $\omega_\cmc = 124^\circ$, and $P_\cmc = 21.5089$ days. For the purposes of this demonstration, we assumed uniform surface brightness of the stellar disk (i.e. no limb darkening) and we enlarged the mass and radius of the moon. Furthermore, we added a synthetic noise model with autocorrelated effects, described by an Autoregressive Integrated Moving Average (ARIMA) model \citep[e.g.][]{wilson}. An $ARIMA(p, d, q)$ process consists of an autogregressive model of the order $p$ and a moving average model of the order $q$, with a $d$-order integration. In time series analysis, integration can be thought of as the inverse of \textit{differencing}, a smoothing operation that generates from the time series $y(t)$ the series $y'(t)$ by representing the change between each consecutive element of $y(t)$, thus:
\begin{equation}
y'_i(t) = y_i(t)-y_{i-1}(t).   
\end{equation}
The autoregressive and moving average processes can be represented as
\begin{align}
    y_i(t) &= c_1 + \sum_{k = 1}^p \phi_k y_{i-k}(t) + \varepsilon_i\\
    y_i(t) &= c_2 + \varepsilon_i + \sum_{k = 1}^q \theta_k \varepsilon_{i-k}.
\end{align}
Here $c_{1,2}$ are constants, $\phi_k$ and $\theta_k$ are the coefficients of the two processes, while the noise terms $\varepsilon$ are drawn from a Gaussian distribution with zero mean (i.e. white noise). In our case (Figure \ref{fig:demo}), we set up the model with the parameters $ARIMA([1, -0.5], [1], [0.5, -1])$\footnote{In this case, $p=q=2$ and $d = 1$. The coefficients $\theta_{1,2}$ and $\phi_{1,2}$ are expressed in the brackets.}. Figure \ref{fig:demo} clearly shows that the algorithm is capable of handling different angular velocities for the planet and its satellite (manifesting in different transit durations for the two objects).


\subsection{A realistic configuration}

\begin{figure*}
    \centering
    \includegraphics[width = \textwidth]{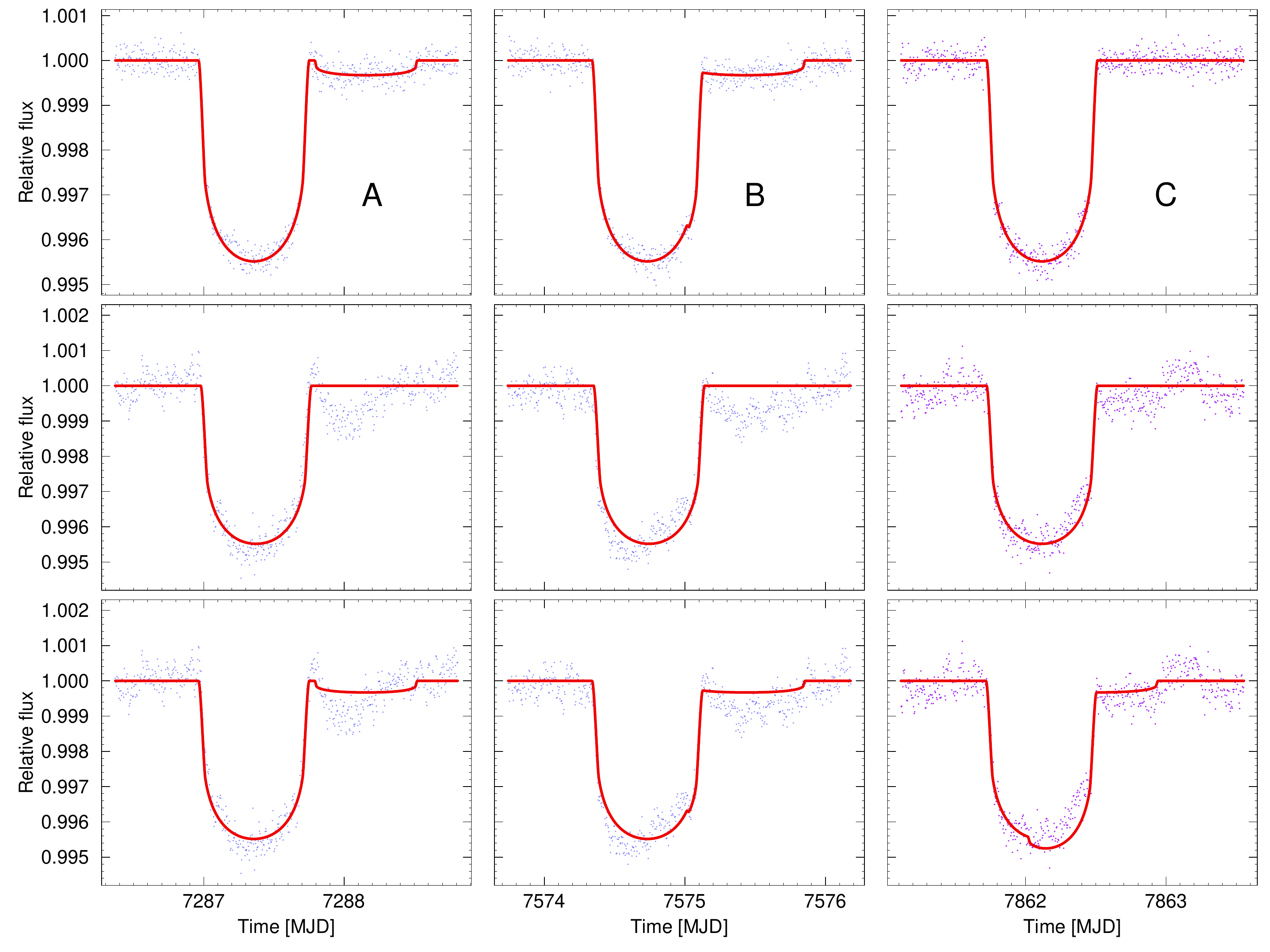}
    \caption{Demonstration of time-correlated noise mimicing transits of an exomoon. In cases A (left column) and B (middle column), the transits of the moon are present in the synthetic photometry (points), case C (right column) presents the transit of an exoplanet only but leads to false detection because of the correlated noise in combination with short observations. The top row represents the ideal case where only white noise is present on the light curve. The middle and bottom rows also have an ARIMA noise model superimposed upon them. The synthetic photometry in the bottom row is the same as in the middle row, but the solid red line represents a planet-and-moon scenario. We emphasise that these figures are not results of light curve fitting.}
    \label{fig:kip}
\end{figure*}

We also explore a realistic planet-moon case. To that end, we adopted the parameters of the planet and its candidate satellite from \cite{2018SciA....4.1784T}, with arbitrary limb darkening parameters, while restricting the orbital plane of the moon to coincide with that of the planet. The left and middle columns in  Fig. \ref{fig:kip} show two transits with arbitrarily positioned moons. In the rightmost column, we demonstrate a false-positive detection in a no-moon case. 

The top panels show the synthetic input transits with a white noise component only (with a $200$ ppm point-to-point scatter, compared to the $\sim270$ ppm transit depth of the moon). The three columns present two systems with moons (A and B) and a moonless case (C). 
The middle row shows the same input signals, but with a red noise component added as well. To be realistic, here we introduced the red noise via an $ARIMA([0.6819], [1], [-0.3603,-0.5747])$ process, cloning the noise of the residuals of the original measurement of \cite{2018SciA....4.1784T}.  The bottom row presents a comparison between the ``moon'' solution and the synthetic planet+moon+red noise.
We observe that in cases A and B the presence of the moon is convincing in the synthetic measurements even by visual inspection. On the other hand, in scenario (C) where the moon is absent, we encounter a false positive detection due to the similarity of the red-noise-induced signal to that of the moon's.

One possible way to exclude false detections like this is to schedule observations that are long enough so that a noise filtering algorithm \citep[e.g. the wavelet-filter of][]{2009ApJ...704...51C} is able to discern the actual noise properties in detail.

The true advantage of this filtering technique lies in the fact that it does not require prior knowledge of the exact source or realisation of the time-correlated noise, since it operates by constructing a time series model in a wavelet basis. It is therefore efficient in removing the red noise down to the time scales of minutes \citep{2021arXiv210811822C}. It is also important to stress that the test carried out with the TLCM thus far (although they were limited to transiting exoplanets) indicate that the radius of the transiting object can be recovered with high precision and accuracy even at low S/N \citep{2021arXiv210811822C, 2022arXiv220801716K}. This would also be one of the most important parameters for a potential exomoon candidate.

Fig.~\ref{fig:kip} has two important lessons for us: 
\begin{itemize}
\item{} When chasing a moon, the measurement of the null signal (noise only) before and after the transit has to be as long as or longer than the part of the presumed signal itself, to enable the algorithms to distinguish between noise and signal.
\item{} The threshold of a secure detection depends on the length of the noise sampling as well as the actual geometrical configuration. This has to be tested even at the observation planning phase, and has to be taken into account in reasoning the feasibility of any kind of measurement in this field.
\end{itemize}
We note however that the possibility of dense rings around Kepler-1625b, or even the as yet unconfirmed Kepler-1625b-i \citep{2022MNRAS.512.1032S} can not be excluded, although exploring that possibility is beyond the scope of this paper and will be the subject of a subsequent one.

\section{Discussion and summary} \label{sec:con}

Motivated primarily by the lack of confirmed exomoons and the work of \cite{2019A&A...624A..95H}, we developed a complex photodynamical model, capable of describing a planet--moon system on an elliptical orbit around a host star in the presence of red noise. We integrated this model into the TLCM code \citep{2020MNRAS.496.4442C}. One of the primary features of the TLCM algorithm is the ability to handle time-correlated noise via the wavelet-based routines of \cite{2009ApJ...704...51C}. 

It is therefore expected that by combining this noise handling method and the photodynamical model presented here, we will be able to ($i$) reject false positive exomoon candidates \citep{2019A&A...624A..95H}, ($ii$) uncover exomoons `hidden' by red noise, and ($iii$) characterise real exomoons with sufficiently high precision and accuracy, as it is commonly done for exoplanets \citep{2021arXiv210811822C}.

The results shown in the paper can be summarised as follows:
\begin{itemize}
\item{} The Exomoon Agent embedded in the TLCM algorithm is introduced. With this development, users can now search for planet+moon solutions in addition to planet-only cases, and will be able to chose the best model family based on the e.g. Akaike and Bayesian Information Criteria.
\item{} As a case study, we investigated possible transits of a planet--moon system with parameters taken from \cite{2018SciA....4.1784T}. We added correlated noise to the simulations using ARIMA clones of the actual residual of the observations.
\item{} We observed that correlated noise components can mimic the characteristic pattern of exomoons, leading to false positive detections. Our results are in accordance with \cite{2019A&A...624A..95H}.
\item{} The single way to avoid these false detections is to take long enough observations. They must include the pre- and post-transit areas of null signal to teach the algorithm the noise properties in details, that can lead to a reliable mitigation. The threshold of a suspected detection depends on the sampling of the observed data, which must be calibrated by accounting for the length of the `red noise only' (or out-of-transit) light curves. 
\item{} With the currently available instruments (including those that are in advanced stages of development such as Ariel), we will likely never be able to claim with full confidence that an exomoon has been detected by modelling a single transit. This is partly because the proper separation of time-correlated noise of astrophysical and instrumental origin itself can be challenging in and of itself, even when the instrumental effects are well understood as in the case of CHEOPS \citep{2020A&A...635A..24H}. In order to separate the temporally correlated noise from the signal of a transit,  the desirable strategy of observing exomoons with HST, JWST or even Ariel  would have to rely on multiple transits, spanning years (due to the relatively long orbital periods of the exomoon-hosting planets). This would, ideally, allow observations of the target system in multiple configurations (Fig. \ref{fig:eclipses}), thus providing a deeper understanding of the probability of false alarm. A strong claim to the presence of an exomoon would have to be the change in occurrence of its transit from observation to observation (relative to the transit of its host). Ideally, TTVs would also have to be observed, and a thorough follow-up would be needed.
\item{}  While an understanding of systematic noise sources is beneficial, the wavelet-based noise handling of TLCM can reasonably be expected to allow us do distinguish between an actual moon and a false detection. This has been the inspiration behind the Exomoon Agent upgrade of TLCM.
\end{itemize}

The software is available publicly for further tests and uses. The thorough validation of the parameter-retrieval capabilities is beyond the scope of this Letter, and it remains in the user's responsibility to look for the stability of solutions in the actual modelling. In an upcoming publication, we will test the parameter stabilities in simulated data sets and data from real measurements. It is expected however, that the Exomoon Agent can be used to reanalyse existing data. 

We expect that the current chase for exomoons will remain with throughout the operations of the current space telescopes. In the search for ever smaller satellites, orbiting planets whose hosts are ever farther away, there will be a clear need for data from larger and more sensitive future space-born telescopes, such as the Large UV Optical Infrared Surveyor \citep[LUVOIR;][]{2019arXiv191206219T}.



\section*{Acknowledgements}

SzK and GyMSz acknowledge the PRODEX Experiment Agreement No. 4000137122 between the ELTE E\"otv\"os Lor\'and University and the European Space Agency (ESA-D/SCI-LE-2021-0025). Project no. C1746651 has been implemented with the support provided by the Ministry of Culture and Innovation of Hungary from the National Research, Development and Innovation Fund, financed under the NVKDP-2021 funding scheme. SzCs gratefully acknowledges the European Space Agency and the PLATO Mission Consortium, whose outstanding efforts have made these results possible. We thank DFG Research Unit 2440: ’Matter Under Planetary Interior Conditions: High Pressure, Planetary, and Plasma Physics’ for support. We also acknowledge support by DFG grants RA 714/14-1,2 within the DFG Schwerpunkt SPP 1992: ’Exploring the Diversity of Extrasolar Planets’. ASE acknowledges support from the Swiss Space Office through the ESA PRODEX program. This work was supported by European Union’s Horizon Europe Framework Programme under the Marie Skłodowska-Curie Actions grant agreement No. 101086149 (EXOWORLD). A.De. acknowledges financial support from the Swiss National Science Foundation (SNSF) for project 200021\_200726. We would like to dedicate this paper to the memory of Prof. István Jankovics (1943-2023), recognizing the profound influence he had on our scientific community and the enduring inspiration he provides to current and future generations of astronomers.

\section*{Data Availability}

No new data were generated or analysed in support of this research. The Transit and Light Curve Modeller software is available at \url{http://transits.hu/}.



\bibliographystyle{mnras}
\bibliography{example} 





\bsp	
\label{lastpage}
\end{document}